\documentclass[showpacs,pra,twocolumn]{revtex4}

\usepackage{amsmath,amssymb}
\usepackage{graphicx}

\begin{document}

\title{Atomic Zitterbewegung}

\author{M. Merkl} 
\affiliation{SUPA, School of Engineering and Physical Sciences, Heriot-Watt University, Edinburgh EH14 4AS, United Kingdom}
\author{F. E. Zimmer}
\affiliation{SUPA, School of Engineering and Physical Sciences, Heriot-Watt University, Edinburgh EH14 4AS, United Kingdom}
\author{G. Juzeli\=unas}
\affiliation{Institute of Theoretical Physics and Astronomy of Vilnius University, A.Gostauto 12, 01108 Vilnius, Lithuania}
\author{P. \"Ohberg}
\affiliation{SUPA, School of Engineering and Physical Sciences, Heriot-Watt University, Edinburgh EH14 4AS, United Kingdom}

\begin{abstract}Ultra-cold atoms which are subject to ultra-relativistic dynamics are investigated. By using optically induced gauge potentials we show that the dynamics of the atoms is governed by a Dirac type equation. To illustrate this we study the trembling motion of the centre of mass for an effective two level system, historically called Zitterbewegung. Its origin is described in detail, where in particular the role of the finite width of the atomic wave packets is seen to induce a damping of both the centre of mass dynamics and the dynamics of  the populations of the two levels. 
\end{abstract}

\maketitle

\section{Introduction}

Ultra-cold atoms obeying ultra-relativistic dynamics would be a match made in heaven. 
An atomic cloud cooled down to nano-Kelvin temperatures  offers an unprecedented opportunity to 
study and manipulate a true quantum gas. Relativistic dynamics, on the other hand, seems at 
first sight incompatible with the concept of an ultra-cold quantum gas.  This is not
 necessarily the case. It has been noted in earlier works on atomic gases \cite{ruostekoski_2002,juzeliunas_2008,vaishnav_2007},  but also in layers of graphene \cite{Tworzydlo-PRL-2006}, that it is indeed possible to study relativistic dynamics for systems which are inherently non-relativistic. Counter-intuitively this is the case in the low-momentum limit. 
 
Already in the early days of quantum mechanics the dynamics of relativistic particles attracted a lot of attention. It was soon realised that a number of counter-intuitive results would follow in the relativistic limit, such as the Zitterbewegung, i.e., the trembling motion
of the centre of mass of a wave packet. Zitterbewegung has attracted a lot of interest over the years and is continuing to be an active field of research which involves a broad range of physical systems \cite{Cserti-PRB-2006,Katsnelson-EPJB-2006,Tworzydlo-PRL-2006,Trauzettel-PRB-2007}.

The notion of Zitterbewegung itself  has its roots in the work of  Schr\"odinger 
on the motion of the free particle  based on Dirac's  relativistic generalization of a wave equation
for spin-1/2 particles  \cite{Schroedinger-SPAW-1930,Dirac-PRS-1928a,Dirac-PRS-1928b}. From these
early days on the existence of the Zitterbewegung for relativistic particles has also been the subject of some controversy \cite{huang_1952,krekora_2004}.  

Zitterbewegung results from the interefence of the positive and negative energy solutions of the
free Dirac equation. The frequency of this interference process is determined by the energy gap
between the two possible solution manifolds.  In case of the historically first discussed motion of
a free electron wave packet this energy gap is on the order of twice the rest energy $2m_e c^2$ of
the electron, {\it i.e.}, the energy necessary to create an
electron-hole pair. For a complete description the concepts of quantum field theory, {\it i.e.}, particle
creation and annihilation, have to be introduced. If the Zitterbewegung is driven by a process which corresponds to an energy of the order of the rest mass, then this unfortunately also makes it rather unlikely to observe the trembling motion with real electrons  \cite{krekora_2004}. 

A number of physical systems, can, however, be described by an effective
Dirac equation. For these systems the creation energy of all participating real particles is much larger than the gap energy and thus the processes of particle creation and annihilation can be disregarded.  Moreover, the considered \textit{particles} are not real but of a \textit{quasi-particle} nature, e.g. spin-states. The combination of these features allow
us to study the phenomenon of Zitterbewegung in regimes far away from its initial discovery -- the motion of a free electron.

In this paper we will study the trembling centre of mass motion of a two-level system which is subject to an off-diagonal matrix gauge potential. In the limit of low momenta and strong gauge fields the dynamics is well described by a Dirac-type equation. In recent papers \cite{jacob_2007,juzeliunas_2008} it has been shown how atoms with an internal tripod level structure, see  Fig. \ref{fig_exp}, may evolve under the influence of an effective non-Abelian vector potential. Here we restrict the motion to only one direction (see also Ref. \cite{vaishnav_2007} for a two-dimensional description of Zitterbewegung).

The paper is organized as follows. In the following section we will briefly outline the derivation of the gauge potentials for the spin system. As an example of the resulting dynamics in the presence of non trivial gauge potentials we study the Zitterbewegung for neutral atoms. Finally we discuss the phenomenon of Rabi-type oscillations occurring in this context and the damping mechanisms due to finite size effects.

%
\section{The equation of motion}
\label{sec:TheModel}
%

In the following we will assume the motion of the atoms to be restricted  to one dimension. We choose 
our coordinate  system such that the x-axis is aligned along that particular dimension. 
A gas of ultracold atoms can be considered dynamically one-dimensional if the corresponding transversal energy scale given by the transversal trapping frequency is much higher than all other energy scales, such as the temperature or chemical potential in the presence of collisional interactions.

\begin{figure}
\includegraphics[width=8cm]{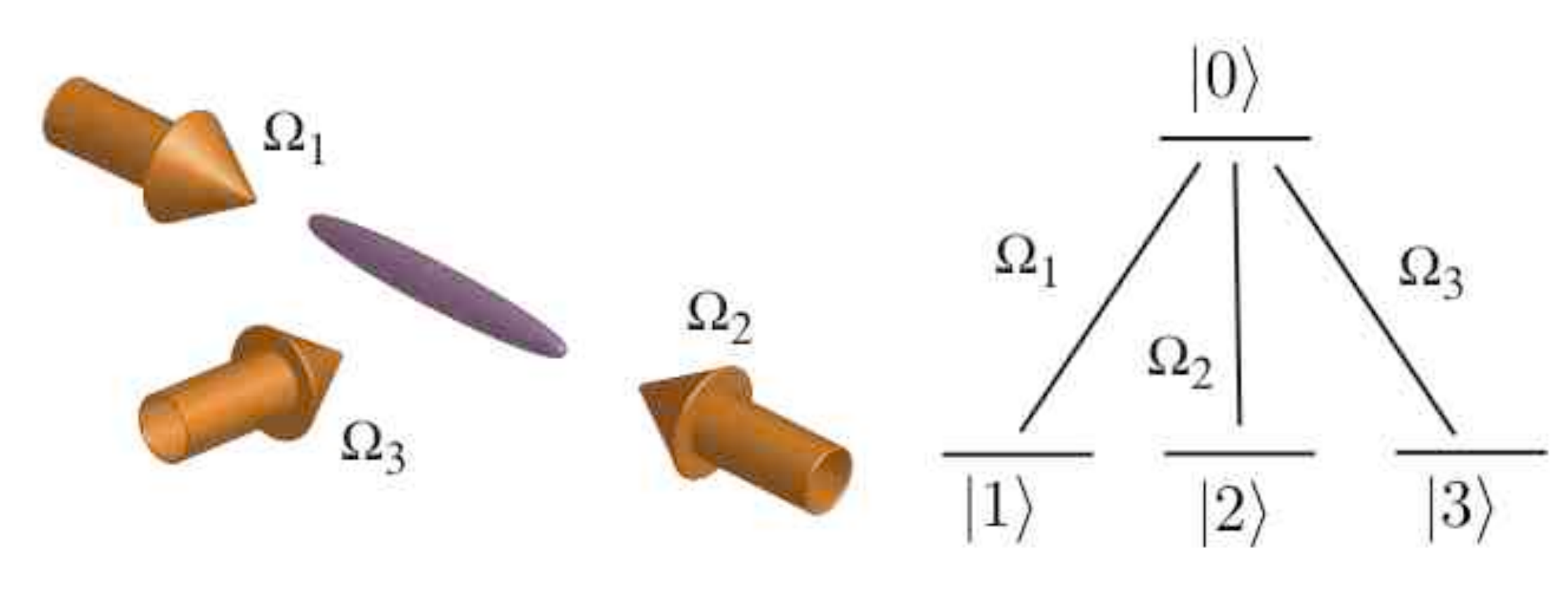}
\caption{One possible laser configuration for the tripod system which results in a non-trivial gauge potential for the two corresponding dark-states.} \label{fig_exp}
\end{figure}

To the effectively one-dimensional cloud of cold atoms we apply the scheme for inducing non-Abelian 
gauge potentials as presented in \cite{Ruseckas_2005d},  and obtain in the limit of low momenta
a {\it quasi-relativistic} situation as shown in \cite{juzeliunas_2008}.
For this purpose, we consider the adiabatic motion of atoms in the presence of three laser beams.
The technique is remarkably versatile and offers the possibility to shape the gauge potentials quite freely. Various possibilities exist for creating non-trivial equations of motions \cite{Ruseckas_2005d}. 
Here we have chosen a laser configuration where 
two of the beams have the same intensity but counter-propagate. The third laser beam
has a different intensity compared to the two other laser beams.  Its wave vector is chosen to be perpendicular to the axis defined by the propagation direction of laser $1$ or $2$. The configuration is depicted in Fig. \ref{fig_exp}.

By defining the total Rabi-frequency $\Omega=\sqrt{\sum_{n=1}^3|\Omega_n|^2}$ and the mixing angle
$\theta$ from $\tan\theta=\sqrt{|\Omega_1|^2+|\Omega_2|^2}/|\Omega_3|$, 
we can write the Rabi
frequencies of the participating laser fields in the following form:
$\Omega_1 = \Omega \sin \theta e^{-i \kappa x} /\sqrt{2}$, 
$\Omega_2 = \Omega \sin \theta e^{i \kappa x} /\sqrt{2} $ and
$\Omega_3 = \Omega \cos \theta e^{-i \kappa y} $.
Applying this notation we find that the interaction Hamiltonian after
the dipole-approximation and in the interaction picture is given by
	\begin{equation} \label{eq hamiltonian interaction1}
	 \hat{H}_{int} = 
	 -\hbar \bigg( \Omega_1 |0\rangle\langle 1|+ \Omega_2 |0 \rangle\langle 2|+\Omega_3 |0 \rangle\langle 3|
 \Bigg) +{h.a.}.
	\end{equation}
in the case of resonant laser fields. 
The Hamiltonian $\hat{H}_{int}$ yields two dark states $|{D_i}\rangle$, $i=1,2$, which contain no contribution from the excited state $|{0}\rangle$:
	\begin{eqnarray} \label{eq def dark state1}
	|{D_1}\rangle &=& \frac{1}{\sqrt{2}} e^{-i \kappa y} \big( e^{i\kappa x}| 1\rangle - e^{-i \kappa x} |2 \rangle \big), \\
	|{D_2}\rangle &=& \frac{1}{\sqrt{2}} e^{-i \kappa y} \cos \theta
	\big( e^{i k x} |1 \rangle - e^{-i k x} |2 \rangle \big) - \sin \theta |3 \rangle.
	\end{eqnarray}
Both dark states are eigenstates of $\hat{H}_{int}$ with zero eigenenergy. They depend on position due to the spatial dependence of the Rabi frequencies $\Omega_i\,$.

The bright state $|{B}\rangle \sim \Omega_1^* |{1}\rangle+ \Omega_2^* |{2}\rangle+ \Omega_3^* |{3}\rangle $ is coupled
to the exited state $|{0}\rangle$ with the Rabi frequency $\Omega$ and therefore separated from the dark states by energies 
$\pm \hbar \Omega$. If $|{\Omega}|$ is large enough compared to any two-photon detuning or Doppler shifts due to the atomic motion, we can
neglect transitions out of the dark states, {\it i.e.}, we use the adiabatic approximation.

In this limit it is sufficient to expand the general state vector $|{\chi}\rangle$ of the quantum system
in the dark state basis
	\begin{equation}\label{eq expand dark states1}
	 |{\chi}\rangle = \sum_{i=1}^2 \Psi_i(\mathbf{r}) |{D_i(\mathbf{r})}\rangle,
	\end{equation}
where $\Psi_i (\mathbf{r})$ are the expansion coefficients. These coefficients represent the centre
of mass motion of the atoms in the dark state $i$. By collecting the wave functions in the
spinor 
	\begin{equation} \label{eq def spinor}
	\bar\Psi = 
	\left( \begin{array}{c} 
	\Psi_1 \\
	\Psi_2 
	\end{array} \right)	
	\end{equation}
we find that the latter obeys the effective Schr\"odinger equation 
	\begin{equation} \label{eq schroedinger equation effectiv1}
	i \hbar \frac{\partial}{\partial t} \bar\Psi = \Big[ \frac{1}{2m} \left(\mathbf{p_x}-\mathbf{\hat A}\right)^2 + \hat{V}+\hat{\Phi} \Big] \bar\Psi,
	\end{equation}
where $\mathbf{p_x}$ denotes the momentum along the x-axis and $m$ is the atomic mass. 
Here $\mathbf{\hat A}$ is an effective vector potential matrix, also called 
the Mead-Berry connection \cite{Mead_1992a,Berry_1984}
and  $\hat{V}$ and $\hat{\Phi}$  are effective scalar potentials matrices. 
The gauge potentials 
$\mathbf{A}_{n,m}=i\hbar \langle{D_n(\mathbf{r})}|{\nabla D_m(\mathbf{r})}\rangle$ 
and 
${\Phi}_{n,m} = \frac{1}{2m} \sum_l \mathbf{A}_{n,l} \mathbf{A}_{l,m}$ 
emerge due to the spatial dependence of the dark states. The other scalar potential 
is defined by $V_{n,m}=\langle{D_n(\mathbf{r})}|\hat{V}|{D_m(\mathbf{r})}\rangle$ with 
$\hat{V}=\sum_{j=1}^3V_j(\mathbf{r})|{j}\rangle\langle{j}|$ and $V_j(\mathbf{r})$ being the 
trapping potential for atoms in the bare state $j$.

With the setup presented in Fig. \ref{fig_exp} these potentials take in $x$-direction the form 
	\begin{eqnarray}
	\label{eq def A}
	\mathbf{\hat A} &=& -\hbar \kappa 
	\left( \begin{array}{cc} 
	0 & \mathbf{e}_x \cos \theta \\
	\mathbf{e}_x \cos \theta & 0 
	\end{array} \right) = -\hbar \kappa' \sigma_x, \\	
	\label{eq def Psip1}
	\hat{\Phi} &=&  \frac{\hbar^2 \kappa^2}{2m}
	\left( \begin{array}{cc} 
	\sin^2 \theta  & 0 \\
	0  &  \sin^2 (2\theta) /4 
	\end{array} \right), \\	
	\label{eq def V1}
	\hat{V} &=& 
	\left( \begin{array}{cc} 
	V_1 & 0\\
	0 & V_1\cos^2 \theta +  V_3 \sin^2 \theta
	\end{array} \right),
	\end{eqnarray}
where we have introduced the notation $\kappa'=\kappa\cos(\theta)$ and assumed that the external trapping potentials for the first two atomic states are the 
same, i.e. $V_1=V_2$. In addition, the external trap in the transversal direction is assumed to dominate over any effective gauge potential in this direction. Hence, we can use an effectively one-dimensional equation of motion. 

\section{The Dirac limit}

The existence of an energy gap is necessary in order to observe 
Zitterbewegung. Such a gap is obtained
by the different, but now constant, trapping potentials $V_1$ and $V_3$ which can be altered
by detuning the corresponding lasers from the atomic transitions. Alternatively, the intensity ratio of the laser beams can be used to adjust the scalar potential.

It is convenient to introduce the following notation
	\begin{equation}
	V_{z} = \frac{1}{2} \Big[{V}_{11} + {\Phi}_{11} 
	- \big({V}_{22} + {\Phi}_{22} \Big) \Big]
	\end{equation}
and shift the zero level of energy. The trapping potential then reads 
	\begin{equation} \label{eq def new trapping}
	\hat{V} + \hat{\Phi} = V_{z} \sigma_z = V_{z}
	\left( \begin{array}{cc}
	1 & 0 \\
	0 & -1
	\end{array} \right).
	\end{equation}
where $\sigma_z$ is one of the Pauli spin matrices. 
In the limit of low momenta, {\it i.e.}, $|p|\ll\hbar\kappa$,  we can 
neglect in Eq. (\ref{eq schroedinger equation effectiv1}) the kinetic energy 
term and are left with an effective Dirac equation 
	\begin{eqnarray}\label{eq eff dirac11}
	i\hbar\partial_t \bar\Psi &=&\left[ -\frac{\mathbf{A}\cdot\mathbf{p_x}}{m}
	+\frac{\mathbf{A}^2}{2m}+V_z \hbar \sigma_z\right]\bar\Psi \\
	&=&\left[ \tilde{c} \sigma_x \mathbf{p}_x
	+V_z \sigma_z\right]\bar\Psi,
	\label{eq eff dirac22}
	\end{eqnarray}
Here $\tilde{c}=\frac{\hbar \kappa}{ m}$ is an effective recoil velocity, which is typically on the order of cm/s for alkali atoms. 
The Dirac limit in Eq. (\ref{eq eff dirac22}) is most clearly justified by considering the corresponding linearised dispersion relation from Eq. (\ref{eq schroedinger equation effectiv1}) \cite{juzeliunas_2008}. Equation (\ref{eq eff dirac22}) is the starting point of our main discussion.
 We note that the $\mathbf{A}^2$ can be absorbed into the 
potential term and we are  hence left with an equation which resembles the Dirac equation for a free relativistic 
particle with the rest energy substituted by the potential energy difference between the two levels.

\begin{figure} 
\includegraphics[width=0.47\textwidth]{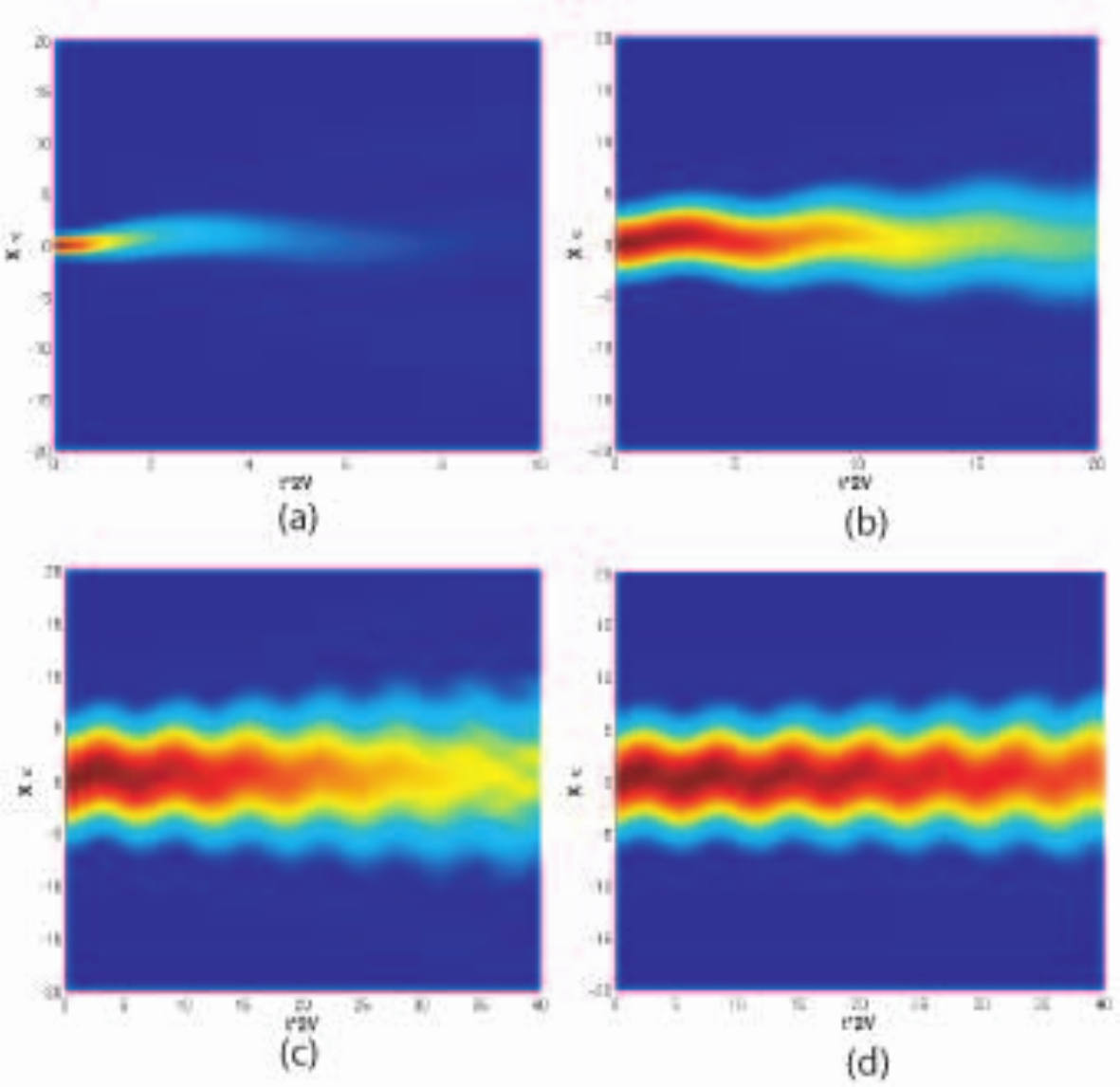}
\caption{Density plot showing the Dirac limit and the role of the initial width of the wave packets. Figures (a)-(c) display the full Schr\"odinger dynamics using eq.(\ref{eq schroedinger equation effectiv1}) for Gaussian initial states with increasing width $\sigma$. This results in a sharper momentum distribution with $p \ll \hbar \kappa$ increasingly fulfilled. The pure Dirac case is shown in figure (d)  for comparison. The dynamics in (a)-(d) shows  in addition to the Zitterbewegung also a damping.
} \label{pic-Dirac-limit}
\end{figure} 

\section{Zitterbewegung}

In most textbooks \cite{Strange, Schwabl_2} Zitterbewegung is derived by solving the Heisenberg equation for the space operator. 
For a free particle with rest mass $m$, the Dirac Hamiltonian $H_{{D}}$ containing the speed of light $c$
	\begin{equation} \label{Dirac equation1}
	H_{D}= c \alpha \mathbf{p} + \beta m c^2
	\end{equation}
is used to obtain the time dependence for the space operator $\mathbf{\hat{x}}$. By using the anticommutation properties of the Dirac $\alpha$ and  $\beta$ matrices one finds
	\begin{eqnarray}\label{eq ZB Dirac1}
	\mathbf{\hat{x}}(t) &=& \mathbf{\hat{x}}(0) + H_{{D}}^{-1}c^2\mathbf{p}t \nonumber\\
	&&\!\!\!\!- \frac{i \hbar c}{2}H_{{D}}^{-1} \Big(\! e^{-2 i H_{{D}} t /\hbar}\!-1\! \Big) \Big(\alpha(0) 
	- c\mathbf{p}H_{{D}}^{-1} \Big) 
	\end{eqnarray}
In equation (\ref{eq ZB Dirac1}) the first and second term describes a motion which is linear in time,
while the third term gives an oscillating contribution, the Zitterbewegung. To observe this trembling motion an initial 4-component spinor state needs to contain positive and negative parts as the $\alpha$ matrix is mixing these. The frequency of the oscillating term can be estimated in the particles rest frame
as $2 m c^2/\hbar$. This typically large energy is the energy difference between a particle and antiparticle. 

\begin{figure} 
\includegraphics[width=0.47\textwidth]{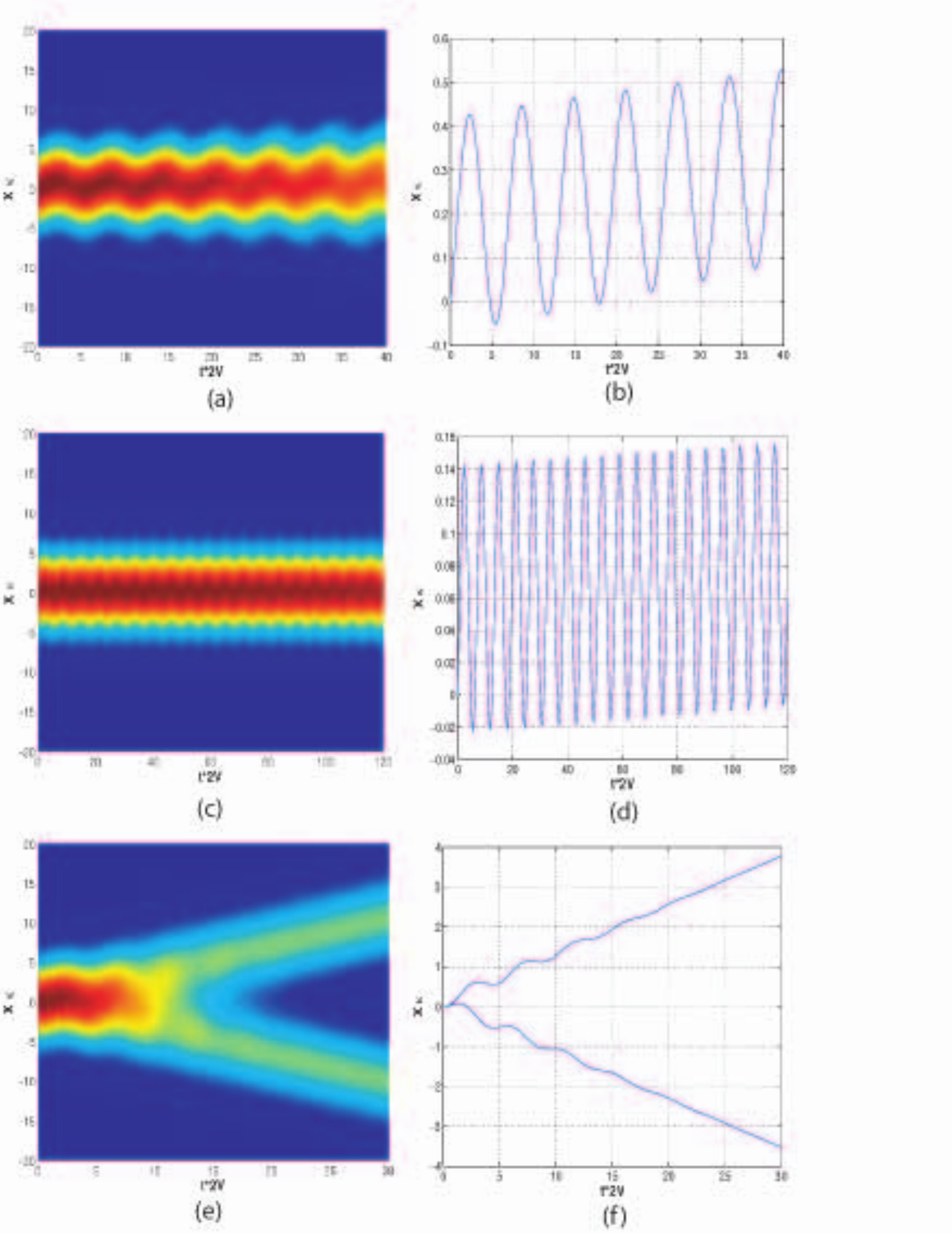}
\caption{Left column: The density as a function of time shows Zitterbewegung for different energy gaps ((a) with $V_z=\frac{\hbar^2\kappa^2}{2m}$ and (c) with $V_z=3\frac{\hbar^2\kappa^2}{2m}$) and in (e) with an initial momentum $k_0=\kappa'$. Right column: The centre of mass shows the expected oscillation ((b) and (d)) with an upward drift. With a initial momentum the behaviour is different, as can be seen in (e) and (f), where the Zitterbewegung breaks down after a few oscillations as the two states are moving in different directions. The initial spinor was in (a)-(d) $(1,1,)^T/\sqrt{2}$ and in (e) and (f) $(1,e^{i\pi/4})^T/\sqrt{2}$.
} \label{pic ZB1}  
\end{figure} 

To obtain Zitterbewegung in cold quantum gases we emphasise that 
the two dark states described by equation (\ref{eq eff dirac22})  are not 
a particle-antiparticle pair, but they are still separated by an energy gap which is generated by the constant 
potential term in Eq. (\ref{eq def new trapping}).
Different to previous work \cite{juzeliunas_2008} the Dirac Hamiltonian now contains an effective rest mass $V_z$. 

\section{Dark state dynamics}

Our system with two degenerate dark states shows not only a relativistic behaviour,
but also properties familiar from two level systems in quantum optics \cite{Barnett}.

In order to see this in more detail we write the spinor $\bar\Psi(x)$ as a combination of slowly varying envelopes, $\phi_i (x)$ $i=1,2$, and coefficients which describe the population of the two dark states,
	\begin{equation} \label{eq rabi ansatz}
	\bar\Psi = \left( \begin{array}{c}
	\phi_1 (x) c_1(t) \\
	\phi_2 (x) c_2(t)
	\end{array} \right).
	\end{equation}
The spatial shape $\phi_i (x)$ 
should change much slower in time than the population $c_i (t)$ of the $i$:th component of the spinor.
The solutions are normalised according to $\langle{\phi_i}|{\phi_i}\rangle=1$ and $|{c_1}|^2+|{c_2}|^2=1$. After inserting the ansatz from equation  (\ref{eq rabi ansatz}) into equation (\ref{eq eff dirac22}) we obtain a set  coupled differential equations for the coefficients:
	\begin{equation} \label{eq differential rabi1}
i\hbar\left( \begin{array}{c}
	\dot c_1 \\
	\dot c_2
	\end{array} \right)=
\left(\begin{array}{c c}
	V_{z1}&\tilde\Omega \\
	\tilde\Omega^*&V_{z2}
	\end{array} \right)\left(
		\begin{array}{c}
	c_1 \\
	c_2
	\end{array} \right)
	\end{equation}
where 
	\begin{equation} 
	\label{aa11}
	\tilde\Omega =  \frac{\tilde{c}}{\hbar}\langle{\phi_2}|{\mathbf p_x}  |{\phi_1}\rangle,\\
	\end{equation}
and
\begin{equation}
	\label{aa22}
	V_{zi} =\langle{\phi_i}| V_z |{\phi_i}\rangle/\hbar.
\end{equation}
The two spin components are coupled, hence the 
solutions to equation (\ref{eq differential rabi1}) will show oscillations between the two dark states with a frequency 
	\begin{equation}\label{eq Rabi frequency1}
	\omega_{R}^2= |{\tilde\Omega}|^2 + \frac{1}{4} \Big( V_{z1}-V_{z2} \Big)^2.
	\end{equation}
For a vanishing overlap integral $\tilde\Omega$, the coupling between the spin components in Eq. (\ref{eq differential rabi1})  becomes zero and we expect no oscillations of the populations. 

If the initial state has a nonzero momentum of the form 
	\begin{equation}
	\phi_i(x)=e^{-x^2/\sigma^2 + i k_0 x}
	\end{equation}
with a width $\sigma$ and momentum $k_0$, we obtain a non-zero $\tilde\Omega$ which will consequently result in population transfer between the two dark states.

\begin{figure} 
\includegraphics[width=0.47\textwidth]{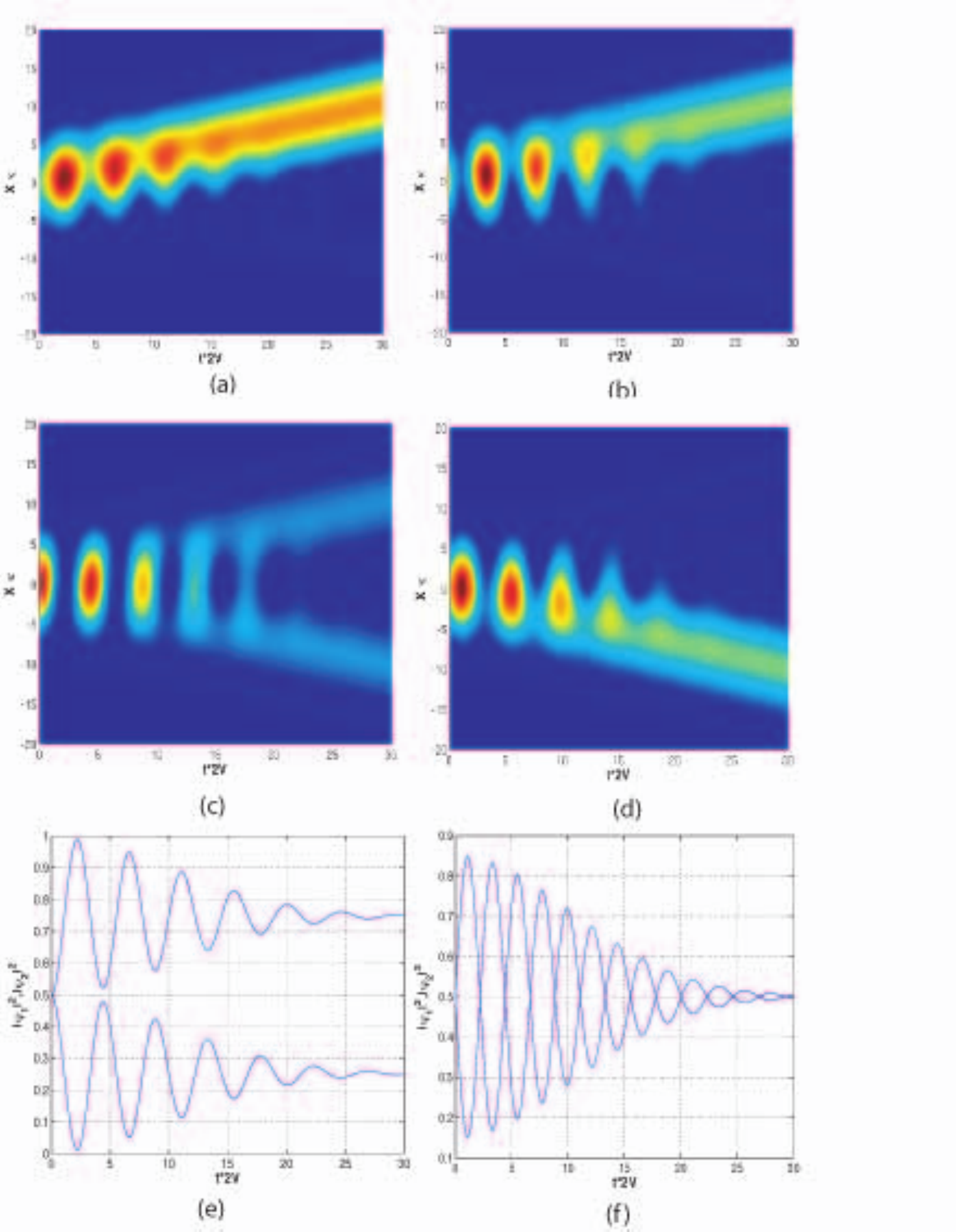}
\caption{ \label{pic Rabi1} Rabi type oscillations superimposing the Zitterbewegung can be seen for an initial preparation with nonzero momentum $k_0=\kappa'$. The dynamics is sensitive to the initial conditions.  Figure (a) and (b) show $|\Psi_1(t,x)|^2$ for the initial spinor spinor $(1,1)^T/\sqrt{2}$ and $(1,e^{i\pi/4})^T/\sqrt{2}$ respectively, whereas figure (c) and (d) show $ |\Psi_2(t,x)|^2$ for the initial spinor $(1,1)^T/\sqrt{2}$ and $(1,e^{i\pi/4})^T/\sqrt{2}$. The population of the dark states is depicted in (e) with $(1,1)^T/\sqrt{2}$ and in (f) with $(1,e^{i\pi/4})^T/\sqrt{2}$.}
\end{figure}
 
\section{Exact solutions in the Schr\"odinger limit}

The Zitterbewegung is typically a transient phenomenon in systems such as graphene and carbon nanotubes \cite{Rusin_2007}. The wave packets corresponding to the positive and negative energies move in different directions which will inevitably cause any interference effect to decay. In the atomic case this is not necessarily so. With the coupled atom-light setup there is no attenuation of the Zitterbewegung for wave packets in the Dirac limit without initial momentum, see Figs. \ref{pic ZB1} (a)-(d). The predicted attenuation \cite{Rusin_2007} is, however, observed if the initial states have a nonzero momentum, as shown in Figs. \ref{pic ZB1}(e) and \ref{pic ZB1}(f).
In this case, an attenuation of the amplitude of the dark state population difference also occures. This is illustrated in Fig. \ref{pic Rabi1}. 

The situation is typically more complicated due to the finite size of the atomic wave packet.
The atomic centre of mass motion is strictly speaking always governed by the Schr\"odinger equation (\ref{eq schroedinger equation effectiv1}), and only in the limit of small momenta compared to the corresponding momentum from the effective gauge potential, are we allowed to construct a Dirac type equation. This, however, would also require us to only use wave packets with zero momentum spread. In the following we will analyse the role of the finite width of the wave packets.

The exact solution of the Schr\"odinger equation (\ref{eq schroedinger equation effectiv1}) with the gauge potentials from Equation (\ref{eq def A})-(\ref{eq def V1}), can readily be written down in momentum space,
	\begin{equation}
	\bar\Psi(k,\tau) = e^{- i(k^2+ 2\sigma_x k + \tilde V_z \sigma_z)\tau} \bar\Psi(k,0),
	\end{equation}
where now the dimensionless $k$ is expressed in units of $\kappa$ and the time $\tau$ in units of $2m/\hbar\kappa^2$. If we choose a Gaussian momentum distribution with a width $\Delta$ for the initial state,
	\begin{equation}\label{eq-initial-gaussian}
	\bar\Psi(k,0)= \frac{1}{\sqrt{\Delta \sqrt{\pi}} } e^{ - (k-k_0)^2/2\Delta^2}  \left( \begin{array}{c} 
	c_1 \\
	c_2 
	\end{array} \right)= \left( \begin{array}{c} 
	\Psi_1(k,0) \\
	\Psi_2(k,0) 
	\end{array} \right),
	\end{equation}
we obtain an exact time dependent solution of the form
\begin{eqnarray} \label{eq spinor kspace t1}
	&&\bar\Psi(k,\tau) =  \frac{1}{\sqrt{\Delta \sqrt{\pi}} } e^{- \frac{(k-k_0)^2}{2\Delta^2} +i(k^2+1)\tau} \nonumber\\
	&&\left( \begin{array}{c}
	c_1 \cos(\omega_k \tau) + \frac{i (c_1 \tilde V_z + c_2 2k ) }{\omega_k} \sin(\omega_k \tau) \\
	c_2 \cos(\omega_k \tau) - \frac{i (c_2 \tilde V_z - c_1  2k ) }{\omega_k} \sin(\omega_k \tau) 
	\end{array} \right),
	\end{eqnarray}
where we have introduced the $k$-dependent frequency
\begin{equation}
\omega_k = \sqrt{4k^2  + \tilde V_z^2}.
\end{equation}
With the solution of $\bar\Psi(k,\tau)$ we can calculate the centre of mass motion for the two-component wave packet where we use the standard definition of the density, $\rho(k,\tau)=|\Psi_1(k,\tau)|^2+|\Psi_2(k,\tau)|^2$, or indeed any dynamics in the populations of the two dark states. 

To illustrate this we choose an initial state with $c_1=c_2=1/\sqrt{2}$. The resulting populations of the two dark states are then given by
\begin{eqnarray}
|\psi_1(\tau)|^2\!\!\! &=&\!\!\! \int_{-\infty}^\infty dk |\Psi_1(0,k)|^2 \big(1 + \frac{2k \tilde V_z }{\omega_k^2} \sin^2(\omega_k \tau) \big)\\
|\psi_2(\tau)|^2\!\!\! &=&\!\!\! \int_{-\infty}^\infty dk |\Psi_2(0,k)|^2 \big(1 -  \frac{2k \tilde V_z }{\omega_k^2} \sin^2(\omega_k \tau) \big).
\end{eqnarray}
The population difference, $\Delta N(t)=|\psi_1(t)|^2-|\psi_2(t)|^2$, can easily be calculated in the limit $\Delta=0$. In this limit the Gaussian initial state turns into a representation of the delta function,
	\begin{eqnarray}
\Delta N(\tau)={\lim_{\Delta\rightarrow 0}}	\int_{-\infty}^\infty dk \frac{4}{ \Delta \sqrt{\pi} }  e^{-\frac{(k-k_0)^2}{\Delta^2}}   \frac{k \tilde V_z}{\omega_k^2} \sin^2(\omega_k \tau) \nonumber \\
	=  \frac{4k_0 \tilde V_z} {\sqrt{4k_0^2 +\tilde V_z^2}} \sin^2(\sqrt{4 k_0^2 + \tilde V_z^2} \tau).
	\end{eqnarray}
From this result we see the importance of the initial momentum $k_0$. For $k_0=0$ there is no transfer of population between the dark states, whereas for a nonzero initial momentum the amplitude of the population oscillation is proportional to $k_0$. In addition, the frequency $\sqrt{4 k_0^2+\tilde V_z^2}$ is $k_0$ dependent as well. 

The expectation value for the centre of mass can be calculated from
\begin{eqnarray} \label{integral-exact-k}
\langle x(\tau) \rangle &=& i\int_{-\infty}^\infty dk \bar\Psi^\dagger(k,\tau) \partial_k \bar\Psi(k,\tau) = \nonumber\\ &&\!\!\!\frac{1}{\Delta \sqrt{ \pi } }\!\! \int_{-\infty}^\infty\!\!\!\!\!\!\! dk  e^{ -\frac{k^2}{\Delta^2} } \Big(\frac{4k^2 }{\omega_k^2} \tau +\frac{\tilde V_z^2 }{\omega_k^3} \sin(2 \omega_k \tau) \Big).
\end{eqnarray}
From the first term under the integral sign we obtain a drift term for the centre of mass,
	\begin{equation}
	x_d=\tau \Big[1 - \sqrt{\pi} \frac{\tilde V_z}{\Delta} e^{ \frac{\tilde V_z^2}{\Delta^2}} {\rm Erfc}(\frac{\tilde V_z}{\Delta})  \Big],
	\end{equation}
where ${\rm Erfc}$ is the complementary Error function. In the limit of $\tilde V_z/\Delta\gg1$, and using the asymptotic expansion of the Error function, 
	\begin{equation}
	{\rm Erfc}(x) = \frac{e^{-x^2}}{x \sqrt{\pi}} \sum_{n}^{\infty} (-1)^n \frac{(2n)!}{n! (2x)^{2n} },
	\end{equation}
we obtain a reduced drift as a function of {\it increasing} $\tilde V_z/\Delta$. This is a finite size effect and stems from the finite width of the wave packet.  

The second term under the integral in (\ref{integral-exact-k}) stands for the Zitterbewegung. Again, in the limit $\tilde V_z/\Delta \gg1$, the integral can be readily calculated and gives for the oscillating part 
\begin{equation}
 x_z=\frac{1}{\tilde V_z} \frac{\sin\Big(2\tilde V_z\tau+\frac{1}{2}\arctan(\frac{\Delta^2}{4\tilde V_z}\tau)\Big)}{(1+\frac{\Delta^4}{16V_z^2}\tau^2)^{1/4}}.
 \label{zb-damp}
\end{equation}
From this expression we see that a spread in the momentum distribution will cause a damping also for the oscillating term of the centre of mass. The damping of the Zitterbewegung is relatively slow, but is inevitable. The underlying equation is after all the Schr\"odinger equation, and only in the limit $\Delta=0$ can we strictly speaking use a Dirac-type equation which results in no expansion of the wave packet, hence no damping. With the full Schr\"odinger equation a free wave packet will always expand, albeit slowly if $\Delta$ is small, and hence will also show a damped Zitterbewegung. 

For a typical alkali atom such as $^{87}$Rb with a wave packet width of $10 \mu m$ one would get ${\Delta^2\tau}/{4\tilde V_z} > 1$ for times larger than $1 ms$, with a centre of mass oscillation frequency of the order of $1 kHz$. Hence a broad wave packet as initial state would favour the detection of the Zitterbewegung. 
\section{Conclusions}
\label{sec:Conclusion}

In this paper we have showed using ultra-cold atoms how Zitterbewegung, known from relativistic physics, is a generic phenomena which will naturally occur in systems with degenerate eigenstates. Interestingly, the atomic scenario offers a number of new possibilities. We are now in  a position to for instance study a system which would correspond to a confined Dirac particle by introducing external atomic potentials, either by optical or magnetical means, for the atoms. This consequently leads us to ponder whether the present system will show Bose-Einstein condensation, and, if so, what will such a quantum state look like \cite{stanesco_2007}. In this context interactions between the atoms will play an important role, where one would be faced with a nonlinear Dirac-type equation to describe the dynamics. The source of the nonlinearity is however nontrivial due to the underlying collisions between the two dark states.

\acknowledgments
This work was supported by the Royal Society of Edinburgh and the UK EPSRC.


\begin{thebibliography}{21}
\expandafter\ifx\csname natexlab\endcsname\relax\def\natexlab#1{#1}\fi
\expandafter\ifx\csname bibnamefont\endcsname\relax
  \def\bibnamefont#1{#1}\fi
\expandafter\ifx\csname bibfnamefont\endcsname\relax
  \def\bibfnamefont#1{#1}\fi
\expandafter\ifx\csname citenamefont\endcsname\relax
  \def\citenamefont#1{#1}\fi
\expandafter\ifx\csname url\endcsname\relax
  \def\url#1{\texttt{#1}}\fi
\expandafter\ifx\csname urlprefix\endcsname\relax\def\urlprefix{URL }\fi
\providecommand{\bibinfo}[2]{#2}
\providecommand{\eprint}[2][]{\url{#2}}

\bibitem[{\citenamefont{Ruostekoski et~al.}(2002)\citenamefont{Ruostekoski,
  Dunne, and Javanainen}}]{ruostekoski_2002}
\bibinfo{author}{\bibfnamefont{J.}~\bibnamefont{Ruostekoski}},
  \bibinfo{author}{\bibfnamefont{G.~V.} \bibnamefont{Dunne}}, \bibnamefont{and}
  \bibinfo{author}{\bibfnamefont{J.}~\bibnamefont{Javanainen}},
  \bibinfo{journal}{Phys. Rev. Lett.} \textbf{\bibinfo{volume}{88}},
  \bibinfo{pages}{180401} (\bibinfo{year}{2002}).

\bibitem[{\citenamefont{Juzeli\=unas et~al.}(2008)\citenamefont{Juzeli\=unas,
  Ruseckas, Lindberg, Santos, and \"Ohberg}}]{juzeliunas_2008}
\bibinfo{author}{\bibfnamefont{G.}~\bibnamefont{Juzeli\=unas}},
  \bibinfo{author}{\bibfnamefont{J.}~\bibnamefont{Ruseckas}},
  \bibinfo{author}{\bibfnamefont{M.}~\bibnamefont{Lindberg}},
  \bibinfo{author}{\bibfnamefont{L.}~\bibnamefont{Santos}}, \bibnamefont{and}
  \bibinfo{author}{\bibfnamefont{P.}~\bibnamefont{\"Ohberg}},
  \bibinfo{journal}{Phys. Rev. A} \textbf{\bibinfo{volume}{77}},
  \bibinfo{pages}{011802} (\bibinfo{year}{2008}).

\bibitem[{\citenamefont{Vaishnav and Clark}()}]{vaishnav_2007}
\bibinfo{author}{\bibfnamefont{J.~Y.} \bibnamefont{Vaishnav}} \bibnamefont{and}
  \bibinfo{author}{\bibfnamefont{C.~W.} \bibnamefont{Clark}},
  \bibinfo{note}{arXiv:0711.3270}.

\bibitem[{\citenamefont{Tworzydlo et~al.}(2006)\citenamefont{Tworzydlo,
  Trauzettel, Titov, Rycerz, and Beenakker}}]{Tworzydlo-PRL-2006}
\bibinfo{author}{\bibfnamefont{J.}~\bibnamefont{Tworzydlo}},
  \bibinfo{author}{\bibfnamefont{B.}~\bibnamefont{Trauzettel}},
  \bibinfo{author}{\bibfnamefont{M.}~\bibnamefont{Titov}},
  \bibinfo{author}{\bibfnamefont{A.}~\bibnamefont{Rycerz}}, \bibnamefont{and}
  \bibinfo{author}{\bibfnamefont{C.~W.~J.} \bibnamefont{Beenakker}},
  \bibinfo{journal}{Phys. Rev. Lett.} \textbf{\bibinfo{volume}{96}},
  \bibinfo{pages}{246802} (\bibinfo{year}{2006}), ISSN
  \bibinfo{issn}{0031-9007}.

\bibitem[{\citenamefont{Cserti and David}(2006)}]{Cserti-PRB-2006}
\bibinfo{author}{\bibfnamefont{J.}~\bibnamefont{Cserti}} \bibnamefont{and}
  \bibinfo{author}{\bibfnamefont{G.}~\bibnamefont{David}},
  \bibinfo{journal}{Phys. Rev. B} \textbf{\bibinfo{volume}{74}},
  \bibinfo{pages}{172305} (\bibinfo{year}{2006}), ISSN
  \bibinfo{issn}{1098-0121}.

\bibitem[{\citenamefont{Katsnelson}(2006)}]{Katsnelson-EPJB-2006}
\bibinfo{author}{\bibfnamefont{M.~I.} \bibnamefont{Katsnelson}},
  \bibinfo{journal}{Eur. Phys. J. B} \textbf{\bibinfo{volume}{51}},
  \bibinfo{pages}{157} (\bibinfo{year}{2006}), ISSN \bibinfo{issn}{1434-6028}.

\bibitem[{\citenamefont{Trauzettel et~al.}(2007)\citenamefont{Trauzettel,
  Blanter, and Morpurgo}}]{Trauzettel-PRB-2007}
\bibinfo{author}{\bibfnamefont{B.}~\bibnamefont{Trauzettel}},
  \bibinfo{author}{\bibfnamefont{Y.~M.} \bibnamefont{Blanter}},
  \bibnamefont{and} \bibinfo{author}{\bibfnamefont{A.~F.}
  \bibnamefont{Morpurgo}}, \bibinfo{journal}{Phys. Rev. B}
  \textbf{\bibinfo{volume}{75}}, \bibinfo{pages}{035305}
  (\bibinfo{year}{2007}), ISSN \bibinfo{issn}{1098-0121}.

\bibitem[{\citenamefont{Schr\"odinger}(1930)}]{Schroedinger-SPAW-1930}
\bibinfo{author}{\bibfnamefont{E.}~\bibnamefont{Schr\"odinger}},
  \bibinfo{journal}{Sitzber. Preu\ss. Akad. Wiss., Phys.-math. Klasse}
  \textbf{\bibinfo{volume}{24}}, \bibinfo{pages}{418} (\bibinfo{year}{1930}).

\bibitem[{\citenamefont{Dirac}(1928{\natexlab{a}})}]{Dirac-PRS-1928a}
\bibinfo{author}{\bibfnamefont{P.~A.~M.} \bibnamefont{Dirac}},
  \bibinfo{journal}{Proc. Roy. Soc. (London)} \textbf{\bibinfo{volume}{A117}},
  \bibinfo{pages}{610} (\bibinfo{year}{1928}{\natexlab{a}}).

\bibitem[{\citenamefont{Dirac}(1928{\natexlab{b}})}]{Dirac-PRS-1928b}
\bibinfo{author}{\bibfnamefont{P.~A.~M.} \bibnamefont{Dirac}},
  \bibinfo{journal}{Proc. Roy. Soc. (London)} \textbf{\bibinfo{volume}{A118}},
  \bibinfo{pages}{351} (\bibinfo{year}{1928}{\natexlab{b}}).

\bibitem[{\citenamefont{Huang}(1952)}]{huang_1952}
\bibinfo{author}{\bibfnamefont{K.}~\bibnamefont{Huang}},
  \bibinfo{journal}{American Journal of Physics} \textbf{\bibinfo{volume}{20}},
  \bibinfo{pages}{479} (\bibinfo{year}{1952}),
  \urlprefix\url{http://link.aip.org/link/?AJP/20/479/1}.

\bibitem[{\citenamefont{Krekora et~al.}(2004)\citenamefont{Krekora, Su, and
  Grobe}}]{krekora_2004}
\bibinfo{author}{\bibfnamefont{P.}~\bibnamefont{Krekora}},
  \bibinfo{author}{\bibfnamefont{Q.}~\bibnamefont{Su}}, \bibnamefont{and}
  \bibinfo{author}{\bibfnamefont{R.}~\bibnamefont{Grobe}},
  \bibinfo{journal}{Phys. Rev. Lett.} \textbf{\bibinfo{volume}{93}},
  \bibinfo{pages}{043004} (\bibinfo{year}{2004}).

\bibitem[{\citenamefont{Jacob et~al.}(2007)\citenamefont{Jacob, \"Ohberg,
  Juzeli\=unas, and Santos}}]{jacob_2007}
\bibinfo{author}{\bibfnamefont{A.}~\bibnamefont{Jacob}},
  \bibinfo{author}{\bibfnamefont{P.}~\bibnamefont{\"Ohberg}},
  \bibinfo{author}{\bibfnamefont{G.}~\bibnamefont{Juzeli\=unas}},
  \bibnamefont{and} \bibinfo{author}{\bibfnamefont{L.}~\bibnamefont{Santos}},
  \bibinfo{journal}{App. Phys. B} \textbf{\bibinfo{volume}{89}},
  \bibinfo{pages}{439} (\bibinfo{year}{2007}).

\bibitem[{\citenamefont{Ruseckas et~al.}(2005)\citenamefont{Ruseckas,
  Juzeliunas, \"Ohberg, and Fleischhauer}}]{Ruseckas_2005d}
\bibinfo{author}{\bibfnamefont{J.}~\bibnamefont{Ruseckas}},
  \bibinfo{author}{\bibfnamefont{G.}~\bibnamefont{Juzeliunas}},
  \bibinfo{author}{\bibfnamefont{P.}~\bibnamefont{\"Ohberg}}, \bibnamefont{and}
  \bibinfo{author}{\bibfnamefont{M.}~\bibnamefont{Fleischhauer}},
  \bibinfo{journal}{Phys. Rev. Lett.} \textbf{\bibinfo{volume}{95}},
  \bibinfo{pages}{010404} (\bibinfo{year}{2005}).

\bibitem[{\citenamefont{Mead}(1992)}]{Mead_1992a}
\bibinfo{author}{\bibfnamefont{C.~A.} \bibnamefont{Mead}},
  \bibinfo{journal}{Rev. Mod. Phys.} \textbf{\bibinfo{volume}{64}},
  \bibinfo{pages}{51} (\bibinfo{year}{1992}).

\bibitem[{\citenamefont{Berry}(1984)}]{Berry_1984}
\bibinfo{author}{\bibfnamefont{M.~V.} \bibnamefont{Berry}},
  \bibinfo{journal}{Proc. R. Soc. A} \textbf{\bibinfo{volume}{392}},
  \bibinfo{pages}{45} (\bibinfo{year}{1984}).

\bibitem[{\citenamefont{Strange}(1998)}]{Strange}
\bibinfo{author}{\bibfnamefont{P.}~\bibnamefont{Strange}},
  \emph{\bibinfo{title}{Relativistic Quantum Mechanics}}
  (\bibinfo{publisher}{Cambridge University Press}, \bibinfo{year}{1998}).

\bibitem[{\citenamefont{Schwabl}(1997)}]{Schwabl_2}
\bibinfo{author}{\bibfnamefont{F.}~\bibnamefont{Schwabl}},
  \emph{\bibinfo{title}{Advanced Quantum Mechanics}}
  (\bibinfo{publisher}{Springer}, \bibinfo{year}{1997}).

\bibitem[{\citenamefont{Barnett and Radmore}(1997)}]{Barnett}
\bibinfo{author}{\bibfnamefont{S.~M.} \bibnamefont{Barnett}} \bibnamefont{and}
  \bibinfo{author}{\bibfnamefont{P.~M.} \bibnamefont{Radmore}},
  \emph{\bibinfo{title}{Methods in Theoretical Quantum Optics}}
  (\bibinfo{publisher}{Oxford University Press}, \bibinfo{year}{1997}).

\bibitem[{\citenamefont{Rusin and Zawadzki}()}]{Rusin_2007}
\bibinfo{author}{\bibfnamefont{T.}~\bibnamefont{Rusin}} \bibnamefont{and}
  \bibinfo{author}{\bibfnamefont{W.}~\bibnamefont{Zawadzki}},
  \bibinfo{note}{arXiv:cond-mat/0702425}.

\bibitem[{\citenamefont{Stanesco and Galitski}()}]{stanesco_2007}
\bibinfo{author}{\bibfnamefont{T.}~\bibnamefont{Stanesco}} \bibnamefont{and}
  \bibinfo{author}{\bibfnamefont{V.}~\bibnamefont{Galitski}},
  \bibinfo{note}{arXiv:0712.2256}.

\end{thebibliography}

\end{document}